# An Unconventional Attempt to Tame Mandelbrot's Grey Swans

Denis M. Filatov[1], Maksim A. Vanyarkho[2]
*Sceptica Scientific Ltd, Carpenter Court, 1 Maple Road, Bramhall, Stockport SK7 2DH, UK*
*{[1]denis.filatov, [2]maksim.vanyarkho}@sceptica.co.uk*

**Abstract.** We suggest an original physical approach to describe the mechanism of market pricing. The core of our approach is to consider pricing at different time scales separately, using independent equations of motion. Such an approach leads to a pricing model that not only allows estimating the volatility of future market prices, but also permits forecasting the direction of the price move. Alongside with that, it is crucial that our model implies no calibration on historical market data. And last but not least, properties of the model's solution are consistent with those of real markets: it has fat tails, possesses scaling and evinces nonlinear market memory. As our model has been derived with the tip of the pen, it may be not a yet another confirmation of the known empirical facts, but a theoretical justification thereto. Tests on real financial instruments prove the competence of our approach.

## 1 Introduction

Due to the works by Benoit Mandelbrot in the early 1960s it is now a common place in the financial community that successive price changes on financial markets are described by probability distributions called fat-tailed [15-17]. A particular class of fat-tailed distributions is Levy-stable, or $\alpha$-stable, probability distributions, where $0 < \alpha \leq 2$. The classical Gaussian distribution corresponds to the limit case $\alpha = 2$ [20]. Besides, Mandelbrot showed successive price changes to be dependent the next of the previous, and such a dependence takes place throughout all time scales, be they minutely, daily, monthly etc. price changes. The discoveries made by Mandelbrot have subsequently been confirmed and extended beyond by other researchers [3, 6, 9].

The presence of fat-tailed probability distributions underlying the mechanism of market pricing is usually disregarded while markets are quiet, but they are recalled each time when a new economic crisis befalls. The effect of fat-tailed probability distributions is observed when there occurs an event (say, an extraordinarily large and unexpected price change) which before was thought to be very improbable. To identify such events, Nassim Taleb suggested the term "black swan" as the contrary to usual, quiet and expected, events – "white swans" [24] (also [7]). More precisely, a "black swan" is the extreme case of the unexpected, the unknown unknown, when the event cannot even be imagined before it happens. The intermediate case, the known unknown, is a "grey swan", when the unexpected event is actually recognisable to happen in theory, but ignored until it appears in practice.

Since the works by Mandelbrot, so far there have been suggested various pricing models aimed to predict the appearance of market "grey swans". As the fat-tailed probability distributions are observed in many physical applications apart from economics, it is natural that many physicists have addressed this problem [2, 5, 8, 10, 12, 14, 18, 21-23, 25-29, 31]. However, at the practical stage econophysical modelling is currently reduced to *imitation* of the pricing process: it is intended to tune the parameters of a mathematical model, typically engaged from some or other area of physics using analogue approach, in such a manner that the result of the subsequent simulation would statistically match the real historical market quotes as much as possible. In other words, a posteriori calibration of the model is required. Obviously, happen a small perturbation in the input data (what, needless to say, takes place every instant in financial markets), a recalibration is required, otherwise the model would stop being able to simulate the real pricing dynamics in the future [4]. In other words, those models possess no forecast strength. Besides, even if to forget for a while of the necessity to calibrate the existing models, the next challenge is how to forecast the sign of future price change. Whereas there have been suggested numerous models allowing to estimate the volatility of future prices, the issue how to distinguish in what direction, up or down, the price will go has not been resolved so far [18, 23, 27].

Apart from the replicative character of existing models, often the model development involves making some or other a priori assumptions either about the markets in general or about a specific behaviour of market agents in particular. In our opinion, such assumptions carry input perturbation risks similar to those mentioned above, in the sense that a tiny debatable supposition underlying an econophysical model may eventually lead to a simulation inadequate to market reality. For instance, we reckon it is irrelevant whether the investors act rationally or irrationally; if the current price has taken into account all available market information or not; etc. Being a complex system, any market has thousands (if not millions) of dependencies between its members and thus various suppositions may or may not be fulfilled [1]. Therefore, efforts should be focused on the development of such a market dynamics model which would reflect the reality without involving disputable suppositions.

In general, we share the key idea of econophysics: economics should not be considered as a qualitatively different science standing apart from physics, and therefore one has to use the methodology of physics for developing economical models. Nevertheless, in order to find out why price changes possess certain properties (including the fat tails and "market memory"), we suggest the opposite way: instead of fitting market reality into a predetermined model we try to develop a model which would be consistent with the reality. In doing so, we use an original quantitative formalisation of the old market concept of two players, the “bull” and the “bear”. As a result, we derive an original equation of motion whose solution bears the properties of fat tails and scaling, as well as evinces nonlinear dependence of successive price changes (“market memory”). Furthermore, as the solution is merely an expression of the future price change subject to the known previous one, it has no a posteriori parameters to tune at all, and hence the model requires no calibration. And yet, due to the asymmetry of the solution in respect of the zero price change the model allows estimating both the volatility and the direction of the future movement.

The paper is organised as follows. In Section 2 we give a qualitative description of our approach. In Section 3 we provide the theoretical results, demonstrating that the properties of the solution to our model are consistent with the properties of real markets which have been empirically established so far beginning from works [15, 16]. Then, in Section 4, we provide results of the numerical experiments with real financial instruments, showing the ability of the model to detect dependencies in successive price changes without preliminary calibration of the model. We also give an analysis of the results. In Section 5 we conclude the paper.

## 2 Physics of Our Approach

We consider the entire set of market agents as an open mechanical system of two players, the “bull” and the “bear” (according to the historically established), each of whom is “tugging” the price to his side. New information thrown into the market is likened to the energy introduced into the system. The energy changes the players’ alignment of forces, thereby changing the price. The concrete material factors raising the alignment of forces are of no importance: those might be bad weather conditions in some or other regions of the planet, changes of the interest rates due to the central banks’ decisions, political instability etc. Obviously, the exact relevant “bullish” and “bearish” factors affecting the price are unknown. Nevertheless, we do not need them. All we have to do is to find out in how many times the probability of a price change is greater or less than the probability of the other, independently of the specific market factors resulting in such a change. This is done via the estimation of the amount of energy introduced into the system: the more the incoming energy, the less likely the corresponding price change will take place. As the system is deemed to be open, the energy introduced can have any value, i.e. it is not required to be conserved over time [4]. We also consider the price to be formed simultaneously at diverse time scales (e.g., minutes, days, months etc.). Enumerating over possible values of the future price change, we determine the least “energy-expensive” and therefore the most probable ones.

A qualitative difference of such an approach from other econophysical techniques consists in that it allows developing a model which takes into account all incoming information, but makes no debatable a priori assumptions about the markets or its particular agents, nor requires a posteriori calibration as it bears no parameters to tune.

## 3 Theoretical Results

Our model is a probabilistic equation of motion of the form

$$\Pr(\Delta x) = \Pr\left({}^{j}\Delta x_{\text{next}} \middle| {}^{j}\Delta x_{\text{prev}}\right)^{1}. \tag{1}$$

For each pair of two successive price changes, ${}^{j}\Delta x_{\text{prev}}$ and ${}^{j}\Delta x_{\text{next}}$, it yields the probability of transition from the first state – the past – to the second one – the future. The model's solution is a probability density function of the future price change subject to the known previous one. (More precisely, the general solution to the model is a family of probability distributions rather than a single density function, but for short we shall be speaking as if there is only one distribution, assuming the previous price change ${}^{j}\Delta x_{\text{prev}}$ known and fixed.)

As the price is being formed at different time scales, there is a separate equation of motion for each scale $j \in \mathbb{N}$. This is an unconventionality of our approach which makes our attempt to solve the old problem – to adequately describe the market pricing mechanism – different from the others. To some extent, the cascade idea may be treated as close [27]. However, as far as we know, the idea to use simultaneously a few equations of motion has not been explicitly formulated so far [18].

The developed model possesses the following remarkable properties:

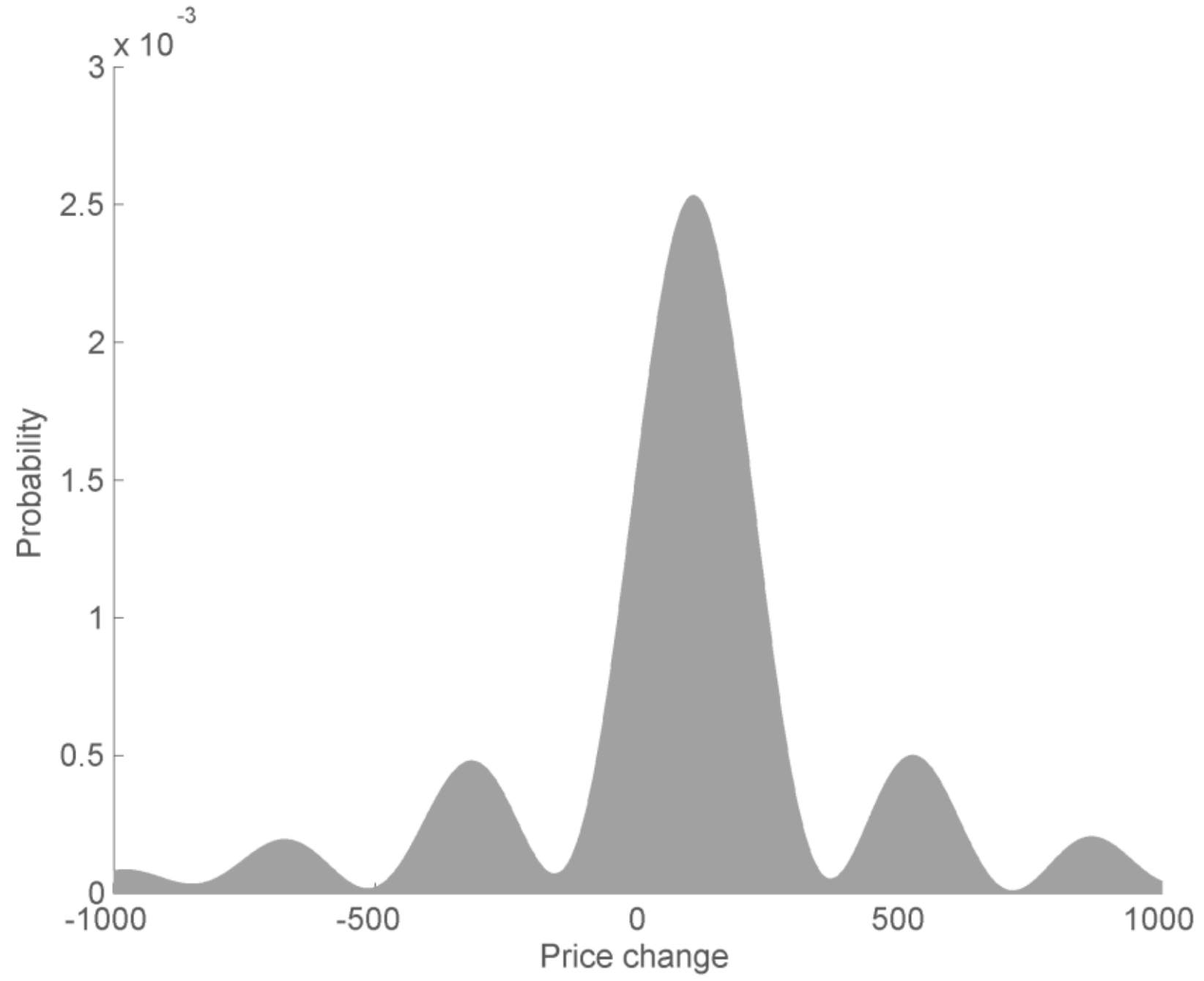

**Fig. 1.** An example of the model's approximate solution – asymmetric fat-tailed probability distribution

1. The probability distribution belongs to the class of asymmetric fat-tailed distributions (Fig. 1). Estimates of the parameter $\alpha$ (the degree of "fat-tailedness") according to [11, 13] yield to be from 0.61 to 0.96, which is essentially far from the Gaussian case $\alpha = 2$.

---

[1] As the explicit expression of the equation of motion is a trade secret of the company, we write it in an implicit manner. However, in Section 3 we demonstrate properties of the model's solution, in Section 4 we provide numerical results obtained with the model, while the model itself is available for independent testing on www.sceptica.co.uk

From the practical standpoint it is crucially important that the solution's asymmetry allows forecasting not only about the absolute value of future price change, but also about the sign of the change, i.e. about the direction thereof.

The wave-like structure of the solution explains the effect of discontinuity of price changes often observed on real markets. In terms of a classical empirical technique – the technical analysis – these are levels of support and resistance that attract the price, so for the price it is more likely to get "nailed" to some or other price level rather than to stop somewhere between the picks [19].

In spite of the presence of nonzero expectation, there is an asymptotic ("tailed") symmetry of the distribution in general, which results in that the correlation coefficient between price changes simulated with the model and taken spaced at a certain time lag decays as the lag grows, rapidly becoming close to zero (Fig. 2). To show this analytically, consider a future quote change ${}^{j}\Delta x_{\text{next}}$ large enough in the absolute value so that

$$\mathrm{r}_{{}^{j}\Delta x_{\text{prev}},{}^{j}\Delta x_{\text{next}}} \sim \sum_{k,l}\left({}^{j}\Delta x_{\text{prev},k} - {}^{j}\overline{\Delta x}_{\text{prev}}\right)\left({}^{j}\Delta x_{\text{next},l} - {}^{j}\overline{\Delta x}_{\text{next}}\right) \to \sum_{k,l}\left({}^{j}\Delta x_{\text{prev},k} - {}^{j}\overline{\Delta x}_{\text{prev}}\right){}^{j}\Delta x_{\text{next},l}. \quad (2)$$

Then, since ${}^{j}\Delta x_{\text{prev},k}$ is fixed, we can rewrite (2) as

$$\mathrm{r}_{{}^{j}\Delta x_{\text{prev}},{}^{j}\Delta x_{\text{next}}} \to \left({}^{j}\Delta x_{\text{prev},k} - {}^{j}\overline{\Delta x}_{\text{prev}}\right)\sum_{l}{}^{j}\Delta x_{\text{next},l}, \quad (3)$$

and hence, due to the asymptotic symmetry, $\sum_{l}{}^{j}\Delta x_{\text{next},l} \to 0$ as $|l| \to \infty$.

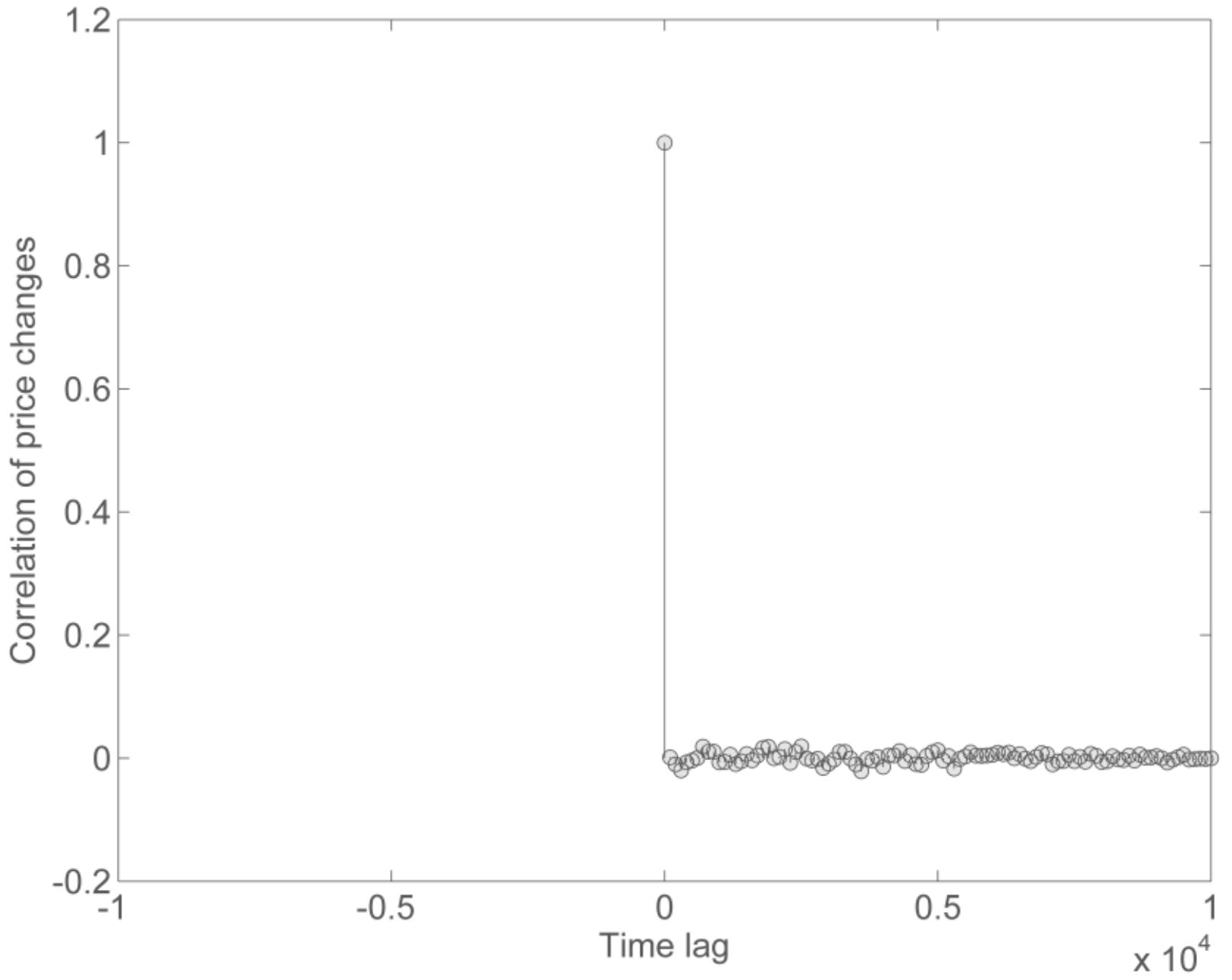

**Fig. 2.** Correlation coefficient between price changes rapidly decays to zero

Nevertheless, the zero correlation does not mean there is no dependence in the price changes – the correlation field simulated with the model does not resemble to be produced by a random walk, as it evinces a certain internal structure (Fig. 3).

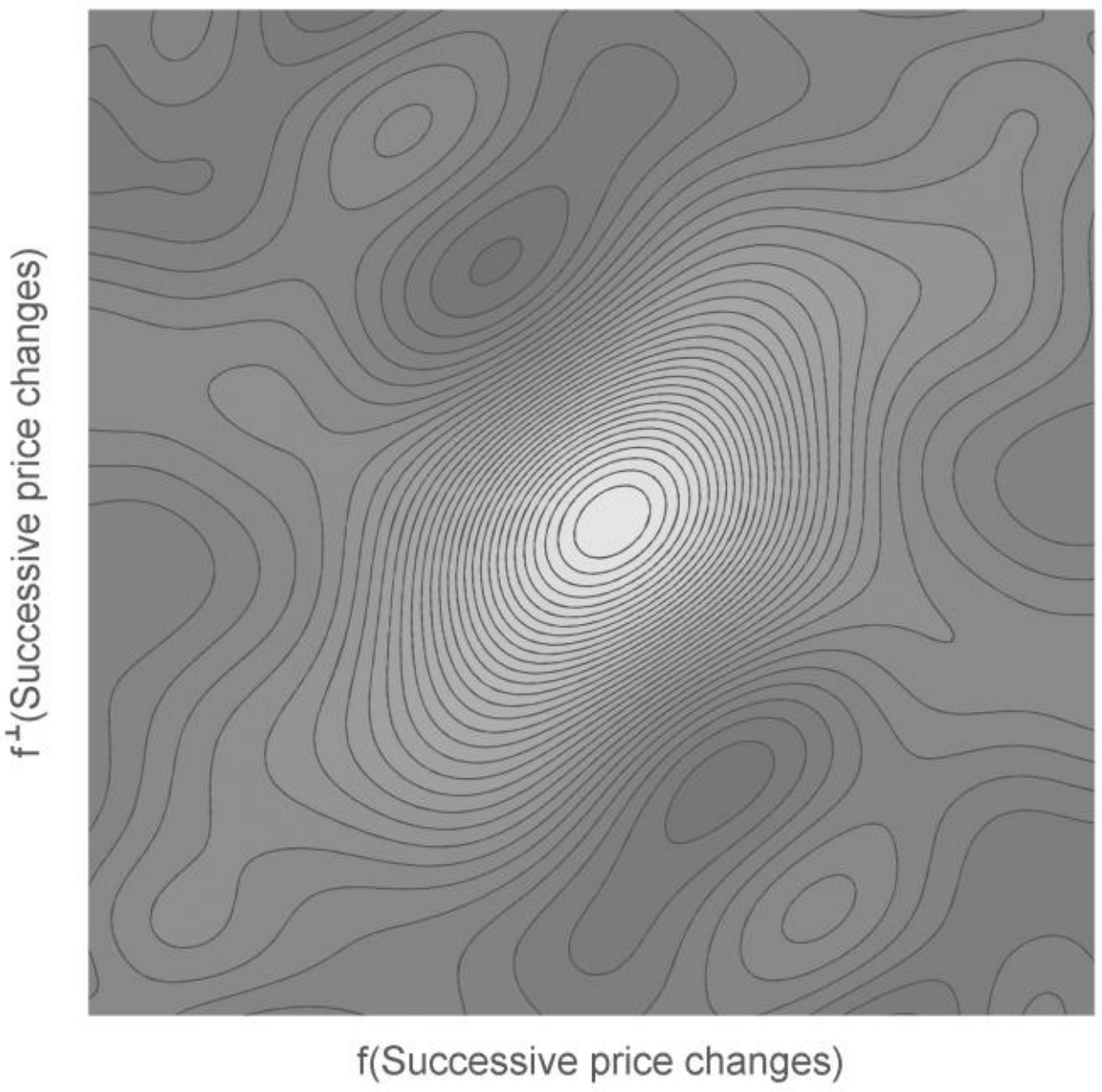

**Fig. 3.** Correlation field (shown smoothed and affinely transformed) clearly indicates to possess a non-random internal structure

Both properties, the zero correlation of price changes and the non-random correlation field, are consistent with the empirical results stating a fast decay of the autocorrelation function when analysing real market data [18, 26], which implies that

2. There exists (actually, *nonlinear*) dependence between price changes.

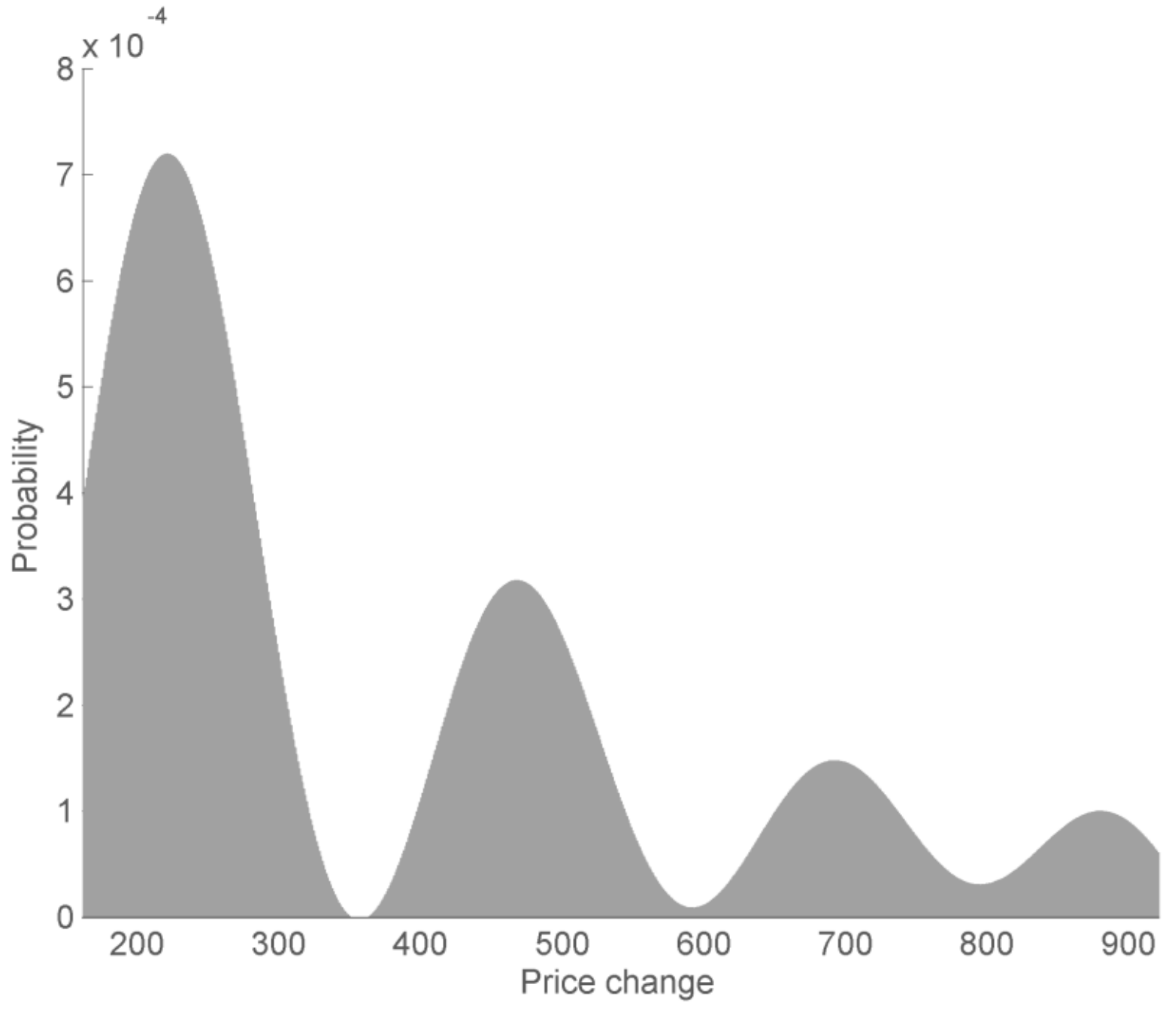

(a)

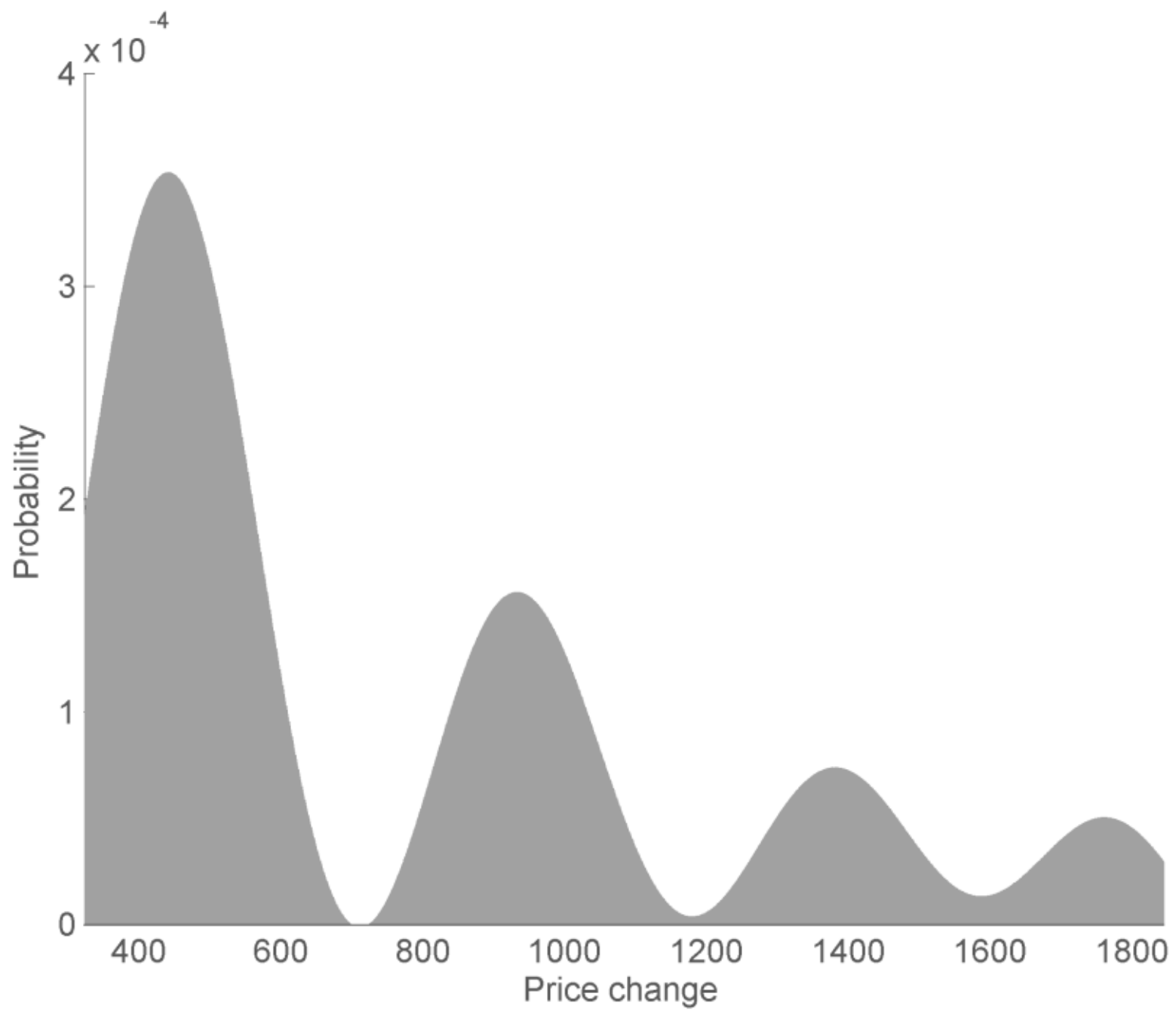

(b)

**Fig. 4.** Solution's tail at two different scales:
the scale of figure (a) is two times less than that of figure (b)

Yet,

3. The distributions' structure is kept unchanged when varying the scale (Fig. 4), which is consistent with the property of scaling of price changes of real market instruments [16, 18].

Finally, the model is also consistent with the real markets' property on volatility [18]: the autocorrelation between squares of price changes is of a power (specifically, linear) law (Fig. 5).

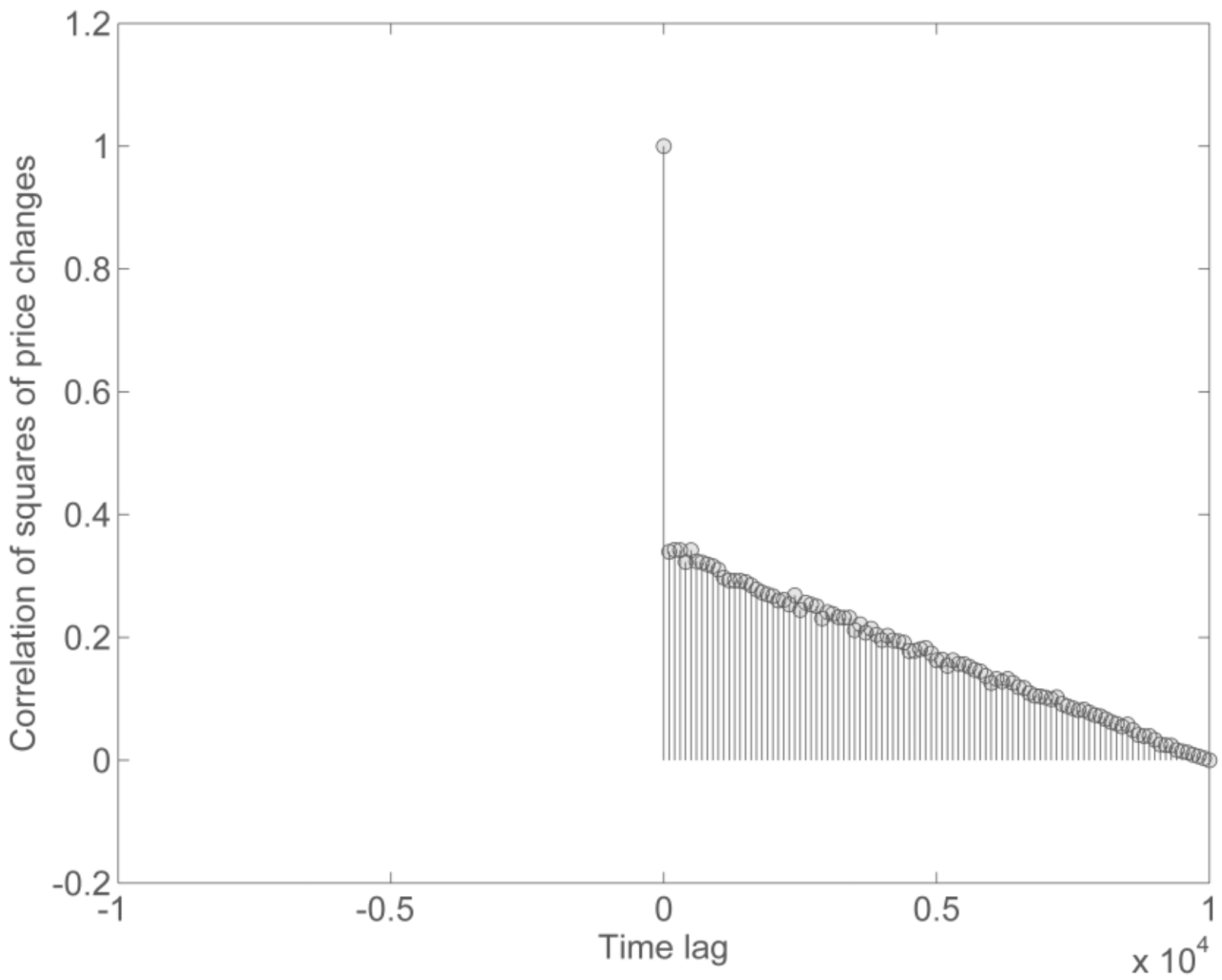

**Fig. 5.** Correlation coefficient between squares of price changes is of a power law

## 4 Numerical Results

To estimate the degree of the model's forecast strength – the ability to give a forecast without preliminary calibration, – we have performed two series of numerical experiments. We considered the model at a fixed time scale $j \in \mathbb{N}$, without scales complexification, as the latter is a nontrivial issue requiring further study.

The main goal was to find out whether the model is able to detect dependencies in successive price changes and predict directions of the movements. Although the model's solution at a fixed ${}^{j}\Delta x_{\text{prev}}$ is a probability distribution, and hence it allows estimating the probability of any specific value ${}^{j}\Delta x_{\text{next}}$, i.e. not only its sign, we carried out the experiments regardless of the future movements' concrete values. Apart from practical importance of the ability to forecast the direction, there were two reasons for that. First, for a model pretending to provide adequate forecast of price changes it is a necessary condition to correctly predict at least the direction, otherwise there is no reason to expect adequacy in forecasting a specific value. Second, whereas the model currently implements no scales complexification, the latter is expected to play an essential role in price formation, and therefore the residual between an actual and predicted price change, computed in the absence of scales complexification, would be useless for subsequent analysis, as it would be impossible to find out if the error is due to the model's principal inadequacy or it is merely because of the neglect of the influence of diverse scales on price formation. Unlike that, the residual between an actual and forecast direction, being computed as a binary function "matched / mismatched", is fairly simple and hence more reliable to analyse.

### *4.1 Series I*

First we tested the model on tick quotes of six real financial instruments as well as on the normal random walk $\mathcal{N}(0,1)$. To compare the results, we also performed tests with randomly permuted original price changes which thereby had assuredly borne no dependence at all.

We took quotes of the currency pair euro-dollar from November 2009 to March 2012, the oil (Brent), the gold and the wheat from 2011 to 2013, as well as the stocks of Google and Apple from September 2011 to December 2013. The mean sample lengths were 351 883, 251 506, 449 731, 67 704, 22 321 and 124 692 successive nonzero price changes respectively [30].

For each nonzero price change $\Delta x_{\text{prev}}$ we computed the probability distribution of the next price change $\Delta x_{\text{next}}$. Having that, we made a probability forecast of the direction of that change and compared it with the real price change's direction that took place on the market. The measurement of residual was quite apparent: if the directions matched, the number of right forecasts was increased by one; if the directions were opposite, we increased the number of wrong forecasts; in the border case – when the real price change was nil – both the numbers of right and wrong forecasts were increased by 1/2. For each month we computed the relative frequency to make the right guess on the direction of future price change, after which we calculated the mean relative frequency over months (Table 1). The supplementary materials provide detailed tables summarising the sample lengths and the corresponding outcomes, as well.

**Table 1.** Relative frequency of success in forecasting the direction of future price change (ticks) for various financial instruments, %

| Financial instrument | Relative frequency's skewness (Pr – 50) | |
|---|---|---|
| | original data | randomly permuted data |
| EUR/USD | **1.32** | 0.00 |
| Oil | **2.26** | -0.00 |
| Gold | **2.84** | -0.01 |
| Wheat | **4.26** | -0.05 |
| Google | **-1.10** | -0.04 |
| Apple | **-1.48** | 0.04 |
| Random walk | 0.02 | -0.04 |

From Table 1 it follows that the chance to forecast the direction for the random walk is foreseeably about 50% (or, in other words, the relative frequency's skewness Pr – 50% is about nil) both on the original and randomly permuted data. However, for the real financial instruments the situation is completely different: while the skewness is still about nil when processing the permuted real data, there is a statistically significant skewness when processing the original real price changes. Note that for Google and Apple the skewnesses are negative, that is the model gives wrong forecasts about the direction more frequently than it provides right predictions. This, however, does not mean the model is inadequate. This merely indicates that for those instruments the influence of major scales is essentially more important than for the others, and hence it should not be disregarded. In other words, accurate forecasting of quote dynamics may significantly imply the use of scales complexification.

### *4.2 Series II*

Apart from the tick quotes, we also performed experiments with the same instruments on minutely (M1, M5, M10, M15, M30), hourly (H1) and daily (D1) quotes with the corresponding closing prices [30]. The time period was since 2000 till 2013 (please, refer to the supplementary materials for details).

In order to obtain stable results on short-sampled non-tick time frames, we used stochastic modelling, performing 100 (on minutes) to 100 000 (on hours and days) repeated simulations for each $\Delta x_{\text{prev}}$. Stochastic modelling has provided more stability to the outcomes, in the sense it has allowed to accurately compute the relative frequency of the appearance of a $\Delta x_{\text{next}}$ subject to the fixed $\Delta x_{\text{prev}}$, so that the errors' dispersions have got smaller.

In Fig. 6 we plot graphs of the skewness of the relative frequency of success for the original and randomly permuted data.

From the figure it follows that forecasting the permuted quote changes (hollow markers) is expectedly a random process – the chance to make the right prediction is about 50%. Specifically, on all the minute time frames there are fairly zero skewnesses with slightly growing dispersions observed when passing to major scales, particularly for Google and Apple. These dispersions are explained by the small numbers of process realisations amplified by rather short samples on the time frames H1 and D1 compared to the minute frames. Empirically estimated, to obtain accurate results a sample should contain not less than 5 000-6 000 entries, while the number of process realisations should be not less than ten.

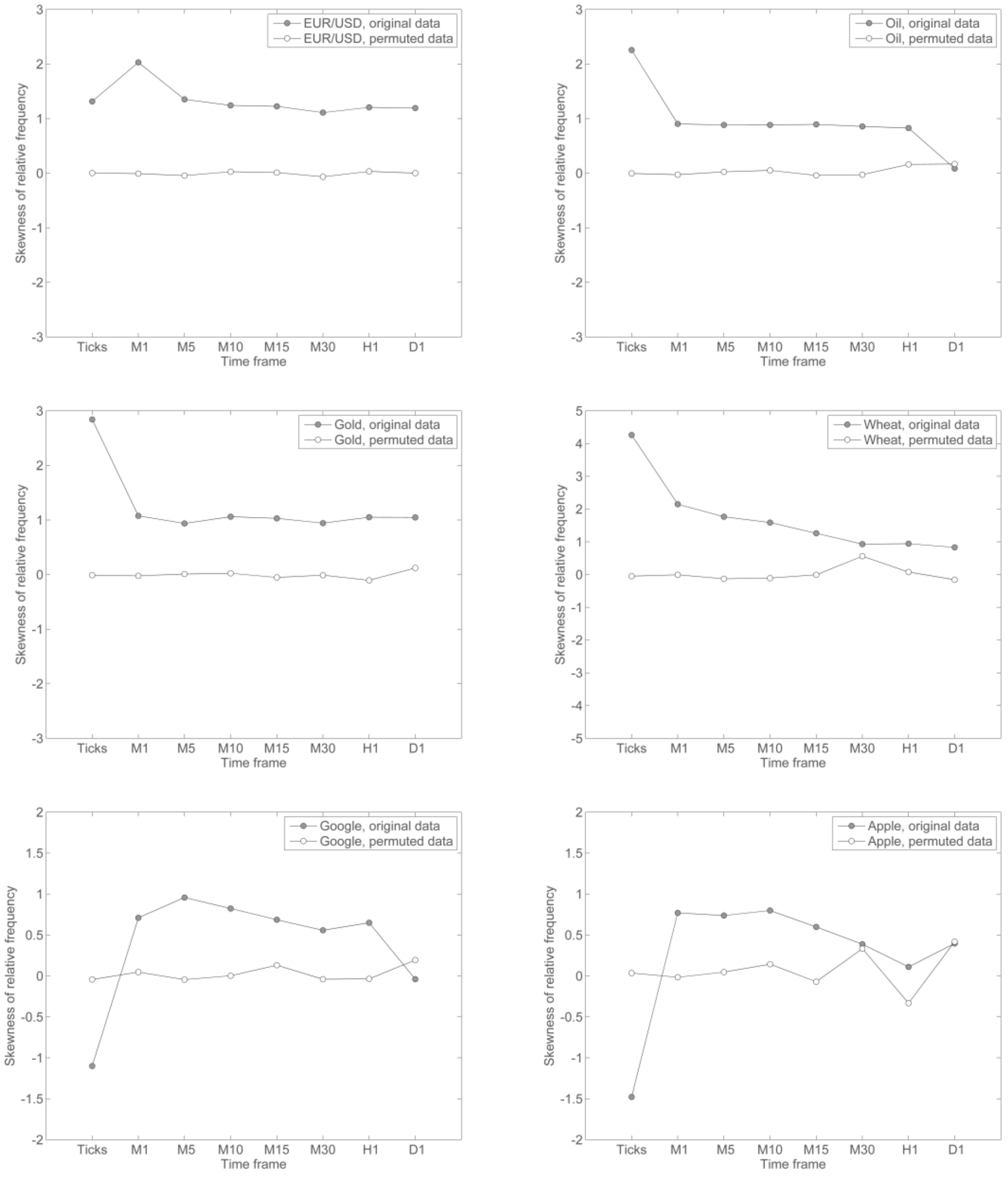

**Fig. 6.** Graphs of the relative frequency's skewness to make the right forecast of the direction of future price change

Nevertheless, forecasting the original data is successful, as statistically significant skewnesses, at the average above 1% (with maximum over 4% on ticks for the wheat), are observed (filled markers). As for some particular values close to zero (Oil on D1, Google on D1, Apple on H1), these are supposed to be because of the lack of scales complexification. Besides, one should not forget that zero skewness (or 50% relative frequency of success, as well) can be an indicator of the purely random market behaviour corresponding to the limit case of the fat-tailed probability distributions – the Gaussian distribution. Whether or not we face it in those particular cases, it does not matter now. But what counts indeed is the fact of the statistically significant relative frequency's skewness observed through nearly all the time frames and on all the instruments, although the model had not been preliminarily calibrated.

Summarising the results of the two series, one can conclude that the model does detect dependence of future price changes on the previous ones having no a priori information about the data being processed.

In the current state the model is available for independent testing on the Internet at `www.sceptica.co.uk`.

## 5 Conclusion

Analysis of the properties of the developed pricing model permits to conclude that the model's solution – a family of fat-tailed probability distributions possessing the property of scaling and detecting nonlinear dependence ("market memory") in successive quote changes – is consistent with the results of the researches on the study of the properties of real market data which have been empirically got so far. But what makes our model qualitatively different from the others is that it is purely analytical, as the probability distribution has resulted as the solution to the corresponding equation of motion containing no a posteriori parameters to tune, i.e. requiring no calibration. In other words, the model consistent with market reality has been derived with the tip of the pen: instead of an *imitation* of the real pricing process it provides a theoretical *justification* to the properties of real market quotes.

For the future research we consider it necessary to find the mechanism of time scales complexification as well as to discover a yet another equation of motion, that which is responsible for the acceleration of trading time [17]. Once the research is completed, we expect those findings, together with the current results, will lead to a pricing model possessing the desired forecast strength suitable for practical purposes.

**EURUSD D1 (Finam)**

| date | original sample length | nonzero sample length | prob. of success | prob. of success (perm.) |
|---|---|---|---|---|
| 2001 | 268 | 258 | 50.7915 | 50.3531 |
| 2002 | 274 | 266 | 51.4157 | 50.5323 |
| 2003 | 309 | 301 | 51.3106 | 51.4351 |
| 2004 | 309 | 307 | 52.0892 | 50.3041 |
| 2005 | 309 | 305 | 50.7288 | 50.7975 |
| 2006 | 311 | 305 | 50.6220 | 48.4397 |
| 2007 | 322 | 314 | 50.6693 | 50.2798 |
| 2008 | 364 | 358 | 51.4076 | 50.4307 |
| 2009 | 363 | 359 | 53.1865 | 49.6569 |
| 2010 | 363 | 357 | 50.5462 | 51.8374 |
| 2011 | 362 | 357 | 50.7220 | 48.0645 |
| 2012 | 359 | 349 | 49.5328 | 48.1974 |
| 2013 | 364 | 359 | 52.4975 | 49.6931 |
| **Mean** | | **323** | **51.1938** | **50.0017** |
| **Skewness** | | | **1.1938** | **0.0017** |
| **Total** | **4277** | **4195** | | |

**EURUSD H1 (Finam)**

| date | original sample length | nonzero sample length | prob. of success | prob. of success (perm.) |
|---|---|---|---|---|
| 2001 | 4673 | 4434 | 51.5835 | 50.7088 |
| 2002 | 3928 | 3755 | 50.8657 | 50.4252 |
| 2003 | 5924 | 5683 | 51.2888 | 49.9907 |
| 2004 | 6056 | 5809 | 51.5153 | 49.8763 |
| 2005 | 6148 | 5880 | 51.1014 | 49.9147 |
| 2006 | 6126 | 5789 | 51.6285 | 50.1072 |
| 2007 | 6095 | 5738 | 51.1639 | 49.7831 |
| 2008 | 6742 | 6229 | 51.0673 | 49.5369 |
| 2009 | 6546 | 6320 | 51.0000 | 50.2965 |
| 2010 | 6503 | 6260 | 50.9889 | 49.5130 |
| 2011 | 6449 | 6218 | 51.0487 | 50.2024 |
| 2012 | 6468 | 6135 | 51.4049 | 50.1878 |
| 2013 | 6460 | 6096 | 51.0300 | 49.8915 |
| **Mean** | | **5719** | **51.2067** | **50.0334** |
| **Skewness** | | | **1.2067** | **0.0334** |
| **Total** | **78118** | **74346** | | |

**EURUSD M30 (Finam)**

| date | original sample length | nonzero sample length | prob. of success | prob. of success (perm.) |
|---|---|---|---|---|
| 2010 | 12963 | 12358 | 50.8098 | 50.0494 |
| 2011 | 12844 | 12279 | 51.0209 | 49.9009 |
| 2012 | 12903 | 12030 | 51.3581 | 49.8906 |
| 2013 | 12848 | 11818 | 51.2527 | 49.8980 |
| **Mean** | | **12121** | **51.1104** | **49.9347** |
| **Skewness** | | | **1.1104** | **-0.0653** |
| **Total** | **51558** | **48485** | | |

**EURUSD M15 (Finam)**

| date | original sample length | nonzero sample length | prob. of success | prob. of success (perm.) |
|---|---|---|---|---|
| 2010 | 25830 | 24183 | 50.9314 | 50.1149 |
| 2011 | 25593 | 23990 | 50.9588 | 50.0384 |
| 2012 | 25753 | 23340 | 51.4139 | 49.8479 |
| 2013 | 25629 | 22885 | 51.5959 | 50.0486 |
| **Mean** | | **23600** | **51.2250** | **50.0125** |
| **Skewness** | | | **1.2250** | **0.0125** |
| **Total** | **102805** | **94398** | | |

**EURUSD M10 (Finam)**

| date | original sample length | nonzero sample length | prob. of success | prob. of success (perm.) |
|---|---|---|---|---|
| 2010 | 38685 | 35678 | 51.0109 | 49.8911 |
| 2011 | 38334 | 35412 | 51.1161 | 49.9412 |
| 2012 | 38595 | 34214 | 51.2968 | 50.2913 |
| 2013 | 38381 | 33527 | 51.5488 | 49.9860 |
| **Mean** | | **34708** | **51.2432** | **50.0274** |
| **Skewness** | | | **1.2432** | **0.0274** |
| **Total** | **153995** | **138831** | | |

**EURUSD M5 (Finam)**

| date | original sample length | nonzero sample length | prob. of success | prob. of success (perm.) |
|---|---|---|---|---|
| 2010 | 77110 | 68928 | 51.0132 | 49.9175 |
| 2011 | 76472 | 68830 | 51.2256 | 49.9751 |
| 2012 | 77079 | 65700 | 51.5193 | 49.9873 |
| 2013 | 76585 | 63816 | 51.6484 | 49.9455 |
| **Mean** | | **66819** | **51.3516** | **49.9564** |
| **Skewness** | | | **1.3516** | **-0.0436** |
| **Total** | **307246** | **267274** | | |

**EURUSD M1 (Finam)**

| date | original sample length | nonzero sample length | prob. of success | prob. of success (perm.) |
|---|---|---|---|---|
| 2010 | 382106 | 306695 | 51.7996 | 50.0032 |
| 2011 | 379862 | 306900 | 51.6605 | 50.0078 |
| 2012 | 382976 | 283656 | 52.2889 | 49.9750 |
| 2013 | 380245 | 267714 | 52.3738 | 49.9845 |
| **Mean** | | **291241** | **52.0307** | **49.9926** |
| **Skewness** | | | **2.0307** | **-0.0074** |
| **Total** | **1525189** | **1164965** | | |

**EURUSD ticks (Dukascopy)**

| date | original sample length | nonzero sample length | prob. of success | prob. of success (perm.) |
|---|---|---|---|---|
| 2009 | 1731989 | 292222 | 51.4771 | 50.0472 |
| | 1646651 | 260935 | 51.5720 | 50.0102 |
| 2010 | 1748179 | 286504 | 51.5981 | 49.9630 |
| | 1958458 | 353742 | 51.3015 | 50.0233 |
| | 2149241 | 339053 | 51.4288 | 49.9966 |
| | 2281966 | 357923 | 51.2236 | 50.0149 |
| | 2192845 | 437322 | 50.9865 | 50.0234 |
| | 2032650 | 338792 | 51.3927 | 50.0335 |
| | 1687019 | 263307 | 51.1433 | 49.9742 |
| | 1746138 | 251492 | 51.0903 | 49.9594 |
| | 1750389 | 241474 | 51.0219 | 50.0292 |
| | 1592282 | 263830 | 50.8041 | 49.9846 |
| | 887503 | 269718 | 51.0741 | 49.9698 |
| | 1373414 | 285660 | 51.1955 | 50.0242 |
| 2011 | 1450299 | 304957 | 51.2258 | 50.0654 |
| | 1497623 | 258403 | 51.4029 | 49.9228 |
| | 1852742 | 269399 | 51.5304 | 50.0711 |
| | 1618076 | 245660 | 51.4661 | 49.9257 |
| | 2289498 | 389464 | 51.2501 | 50.0060 |
| | 2227295 | 369212 | 51.1940 | 50.0391 |
| | 2187642 | 369077 | 51.2710 | 50.0569 |
| | 2699800 | 465729 | 51.2148 | 50.0076 |
| | 2579145 | 433915 | 51.1258 | 49.9553 |
| | 2589607 | 442571 | 51.3047 | 50.0285 |
| | 2616463 | 460975 | 51.5108 | 50.0089 |
| | 2135884 | 308329 | 51.8125 | 49.9760 |
| 2012 | 7408936 | 941166 | 51.8965 | 49.9667 |
| **Mean** | | **351883** | **51.3154** | **50.0031** |
| **Skewness** | | | **1.3154** | **0.0031** |
| **Total** | **57931734** | **9500831** | | |

**Oil Brent D1 (Finam)**

| date | original sample length | nonzero sample length | prob. of success | prob. of success (perm.) |
|---|---|---|---|---|
| 2000 | 366 | 251 | 50.6054 | 49.9630 |
| 2001 | 365 | 252 | 49.8224 | 49.1493 |
| 2002 | 365 | 251 | 50.4591 | 49.3840 |
| 2003 | 365 | 250 | 49.5751 | 50.3993 |
| 2004 | 366 | 257 | 49.0229 | 51.0891 |
| 2005 | 365 | 256 | 49.7901 | 51.1028 |
| 2006 | 365 | 254 | 49.4058 | 50.6726 |
| 2007 | 293 | 257 | 47.9348 | 51.3295 |
| 2008 | 304 | 302 | 51.0186 | 50.3606 |
| 2009 | 304 | 301 | 50.0120 | 50.0756 |
| 2010 | 310 | 309 | 49.1609 | 49.0329 |
| 2011 | 302 | 301 | 51.7178 | 48.8144 |
| 2012 | 309 | 308 | 51.2068 | 51.1529 |
| 2013 | 309 | 306 | 51.4592 | 49.8487 |
| **Mean** | | **275** | **50.0851** | **50.1696** |
| **Skewness** | | | **0.0851** | **0.1696** |
| **Total** | **4688** | **3855** | | |

**Oil Brent H1 (Finam)**

| date | original sample length | nonzero sample length | prob. of success | prob. of success (perm.) |
|---|---|---|---|---|
| 2007 | 1207 | 1162 | 51.5231 | 50.1823 |
| 2008 | 5408 | 5304 | 50.6759 | 50.3451 |
| 2009 | 5628 | 5470 | 50.8149 | 49.9324 |
| 2010 | 5718 | 5579 | 50.7711 | 50.0646 |
| 2011 | 5691 | 5604 | 50.5576 | 50.3336 |
| 2012 | 5745 | 5613 | 50.7186 | 49.9413 |
| 2013 | 5759 | 5594 | 50.7302 | 50.3318 |
| **Mean** | | **4904** | **50.8273** | **50.1616** |
| **Skewness** | | | **0.8273** | **0.1616** |
| **Total** | **35156** | **34326** | | |

**Oil Brent M30 (Finam)**

| date | original sample length | nonzero sample length | prob. of success | prob. of success (perm.) |
|---|---|---|---|---|
| 2010 | 11235 | 10844 | 50.6937 | 50.0752 |
| 2011 | 11265 | 10967 | 50.7466 | 49.8197 |
| 2012 | 11406 | 11039 | 50.9533 | 50.2359 |
| 2013 | 11436 | 10927 | 51.0359 | 49.7631 |
| **Mean** | | **10944** | **50.8574** | **49.9735** |
| **Skewness** | | | **0.8574** | **-0.0265** |
| **Total** | **45342** | **43777** | | |

**Oil Brent M15 (Finam)**

| date | original sample length | nonzero sample length | prob. of success | prob. of success (perm.) |
|---|---|---|---|---|
| 2010 | 21964 | 20797 | 50.7866 | 49.9530 |
| 2011 | 22220 | 21383 | 50.7713 | 50.0836 |
| 2012 | 22473 | 21364 | 50.8618 | 49.8214 |
| 2013 | 22515 | 21150 | 51.1630 | 49.9827 |
| **Mean** | | **21174** | **50.8957** | **49.9602** |
| **Skewness** | | | **0.8957** | **-0.0398** |
| **Total** | **89172** | **84694** | | |

**Oil Brent M10 (Finam)**

| date | original sample length | nonzero sample length | prob. of success | prob. of success (perm.) |
|---|---|---|---|---|
| 2010 | 32310 | 30252 | 50.9873 | 50.0001 |
| 2011 | 32988 | 31428 | 50.7472 | 49.9735 |
| 2012 | 33220 | 31323 | 50.9083 | 50.1675 |
| 2013 | 33253 | 30793 | 50.8912 | 50.0672 |
| **Mean** | | **30949** | **50.8835** | **50.0521** |
| **Skewness** | | | **0.8835** | **0.0521** |
| **Total** | **131771** | **123796** | | |

**Oil Brent M5 (Finam)**

| date | original sample length | nonzero sample length | prob. of success | prob. of success (perm.) |
|---|---|---|---|---|
| 2010 | 60988 | 55733 | 50.7921 | 50.0658 |
| 2011 | 63876 | 59742 | 50.8056 | 49.9833 |
| 2012 | 63780 | 58869 | 50.9904 | 50.0673 |
| 2013 | 63501 | 57428 | 50.9483 | 49.9830 |
| **Mean** | | **57943** | **50.8841** | **50.0249** |
| **Skewness** | | | **0.8841** | **0.0249** |
| **Total** | **252145** | **231772** | | |

**Oil Brent M1 (Finam)**

| date | original sample length | nonzero sample length | prob. of success | prob. of success (perm.) |
|---|---|---|---|---|
| 2010 | 246310 | 208268 | 50.8856 | 50.0088 |
| 2011 | 269982 | 237105 | 50.9128 | 50.0022 |
| 2012 | 262754 | 226990 | 50.8362 | 49.9191 |
| 2013 | 259125 | 216902 | 50.9789 | 49.9636 |
| **Mean** | | **222316** | **50.9034** | **49.9734** |
| **Skewness** | | | **0.9034** | **-0.0266** |
| **Total** | **1038171** | **889265** | | |

**Oil Brent ticks (Finam)**

| date | original sample length | nonzero sample length | prob. of success | prob. of success (perm.) |
|---|---|---|---|---|
| 2011 | 307995 | 207124 | 52.5323 | 49.8346 |
| | 391004 | 275195 | 52.6719 | 50.0184 |
| | 478416 | 339651 | 52.5878 | 49.8824 |
| | 358196 | 252004 | 52.5135 | 50.1375 |
| | 491184 | 368207 | 52.2195 | 49.8714 |
| | 504339 | 365618 | 52.6959 | 50.0187 |
| | 411888 | 294483 | 52.4818 | 50.0557 |
| | 521936 | 390987 | 52.3934 | 50.0460 |
| | 478531 | 356075 | 52.1380 | 50.0424 |
| | 528603 | 386389 | 52.0015 | 49.9867 |
| | 490999 | 358301 | 52.2286 | 49.9746 |
| | 340615 | 241846 | 51.9366 | 49.9963 |
| 2012 | 412974 | 288331 | 52.2306 | 49.9816 |
| | 406220 | 281216 | 52.2611 | 50.0683 |
| | 394789 | 277306 | 52.5236 | 50.0427 |
| | 301998 | 205813 | 52.2639 | 49.8992 |
| | 327351 | 223412 | 51.8927 | 50.0394 |
| | 345973 | 244792 | 52.0093 | 49.9978 |
| | 343146 | 243027 | 51.9004 | 50.0222 |
| | 312626 | 217334 | 52.2074 | 50.1003 |
| | 321138 | 224414 | 52.1628 | 49.9082 |
| | 355862 | 248221 | 52.2919 | 50.1607 |
| | 310138 | 215916 | 52.1474 | 49.9449 |
| | 219937 | 145974 | 51.8976 | 49.9075 |
| 2013 | 330964 | 211578 | 52.6364 | 49.9837 |
| | 301578 | 194041 | 52.5021 | 50.0559 |
| | 266071 | 171395 | 52.4601 | 49.9796 |
| | 379589 | 258009 | 52.2005 | 50.0641 |
| | 300657 | 207287 | 51.9121 | 49.9267 |
| | 299249 | 201231 | 52.2559 | 49.9799 |
| | 320272 | 201733 | 52.2607 | 49.8404 |
| | 294462 | 192631 | 52.1668 | 50.0755 |
| | 297036 | 197164 | 52.0724 | 49.9762 |
| | 346928 | 229961 | 52.2954 | 49.9637 |
| | 302032 | 195803 | 52.3355 | 50.1052 |
| | 215691 | 141744 | 51.9539 | 49.9619 |
| **Mean** | | **251506** | **52.2567** | **49.9958** |
| **Skewness** | | | **2.2567** | **-0.0042** |
| **Total** | **13010387** | **9054213** | | |

**Gold D1 (Finam)**

| date | original sample length | nonzero sample length | prob. of success | prob. of success (perm.) |
|---|---|---|---|---|
| 2000 | 247 | 240 | 50.7284 | 48.9110 |
| 2001 | 247 | 241 | 50.2224 | 50.9538 |
| 2002 | 249 | 247 | 51.5941 | 49.4252 |
| 2003 | 248 | 248 | 52.1535 | 50.2115 |
| 2004 | 244 | 241 | 52.1786 | 49.3311 |
| 2005 | 249 | 241 | 49.1186 | 51.4146 |
| 2006 | 249 | 247 | 50.8623 | 50.3240 |
| 2007 | 271 | 268 | 52.1671 | 52.0107 |
| 2008 | 312 | 308 | 49.3945 | 48.5885 |
| 2009 | 307 | 307 | 50.6380 | 48.6783 |
| 2010 | 307 | 306 | 51.8117 | 50.2367 |
| 2011 | 306 | 305 | 50.3546 | 51.1520 |
| 2012 | 312 | 311 | 51.3666 | 50.1369 |
| 2013 | 312 | 312 | 52.0775 | 50.3168 |
| **Mean** | | **273** | **51.0477** | **50.1208** |
| **Skewness** | | | **1.0477** | **0.1208** |
| **Total** | **3860** | **3822** | | |

**Gold H1 (Finam)**

| date | original sample length | nonzero sample length | prob. of success | prob. of success (perm.) |
|---|---|---|---|---|
| 2007 | 1840 | 1783 | 50.2736 | 49.4310 |
| 2008 | 6011 | 5892 | 50.8999 | 50.0530 |
| 2009 | 6106 | 5937 | 51.4324 | 49.6425 |
| 2010 | 6124 | 5944 | 51.8022 | 50.0999 |
| 2011 | 6107 | 5987 | 51.1233 | 50.1665 |
| 2012 | 6147 | 6026 | 51.0932 | 49.9841 |
| 2013 | 6160 | 6005 | 50.7275 | 49.8897 |
| **Mean** | | **5368** | **51.0503** | **49.8952** |
| **Skewness** | | | **1.0503** | **-0.1048** |
| **Total** | **38495** | **37574** | | |

**Gold M30 (Finam)**

| date | original sample length | nonzero sample length | prob. of success | prob. of success (perm.) |
|---|---|---|---|---|
| 2010 | 11996 | 11524 | 51.1910 | 50.0680 |
| 2011 | 11975 | 11661 | 50.6694 | 49.9577 |
| 2012 | 12036 | 11624 | 51.0310 | 49.8772 |
| 2013 | 12061 | 11650 | 50.8760 | 50.0537 |
| **Mean** | | **11615** | **50.9419** | **49.9892** |
| **Skewness** | | | **0.9419** | **-0.0109** |
| **Total** | **48068** | **46459** | | |

**Gold M15 (Finam)**

| date | original sample length | nonzero sample length | prob. of success | prob. of success (perm.) |
|---|---|---|---|---|
| 2010 | 23869 | 22608 | 51.2647 | 49.8656 |
| 2011 | 23712 | 22816 | 51.0798 | 49.8917 |
| 2012 | 23866 | 22778 | 50.9130 | 49.9857 |
| 2013 | 23905 | 22794 | 50.8644 | 50.0466 |
| **Mean** | | **22749** | **51.0305** | **49.9474** |
| **Skewness** | | | **1.0305** | **-0.0526** |
| **Total** | **95352** | **90996** | | |

**Gold M10 (Finam)**

| date | original sample length | nonzero sample length | prob. of success | prob. of success (perm.) |
|---|---|---|---|---|
| 2010 | 35659 | 33266 | 51.1829 | 50.0066 |
| 2011 | 35535 | 33903 | 50.9633 | 50.0010 |
| 2012 | 35794 | 33759 | 51.0554 | 50.0423 |
| 2013 | 35849 | 33780 | 51.0402 | 50.0439 |
| **Mean** | | **33677** | **51.0605** | **50.0235** |
| **Skewness** | | | **1.0605** | **0.0234** |
| **Total** | **142837** | **134708** | | |

**Gold M5 (Finam)**

| date | original sample length | nonzero sample length | prob. of success | prob. of success (perm.) |
|---|---|---|---|---|
| 2010 | 70942 | 64425 | 50.9757 | 50.0316 |
| 2011 | 70625 | 66042 | 50.8580 | 50.0091 |
| 2012 | 71236 | 65719 | 50.9197 | 50.0248 |
| 2013 | 71291 | 65700 | 50.9909 | 49.9751 |
| **Mean** | | **65472** | **50.9361** | **50.0102** |
| **Skewness** | | | **0.9361** | **0.0101** |
| **Total** | **284094** | **261886** | | |

**Gold M1 (Finam)**

| date | original sample length | nonzero sample length | prob. of success | prob. of success (perm.) |
|---|---|---|---|---|
| 2010 | 336088 | 274998 | 51.1945 | 49.9739 |
| 2011 | 342437 | 295045 | 50.9563 | 49.9489 |
| 2012 | 345817 | 289821 | 51.0719 | 50.0053 |
| 2013 | 346408 | 290101 | 51.0821 | 49.9777 |
| **Mean** | | **287491** | **51.0762** | **49.9765** |
| **Skewness** | | | **1.0762** | **-0.0236** |
| **Total** | **1370750** | **1149965** | | |

**Gold ticks (Finam)**

| date | original sample length | nonzero sample length | prob. of success | prob. of success (perm.) |
|---|---|---|---|---|
| 2011 | 582877 | 370902 | 53.3471 | 50.0372 |
| | 503449 | 315735 | 53.1343 | 49.8307 |
| | 561233 | 363860 | 53.0942 | 50.0857 |
| | 564048 | 365062 | 53.2326 | 49.9412 |
| | 714774 | 499956 | 52.7888 | 50.0300 |
| | 572704 | 368042 | 53.1075 | 50.0416 |
| | 541818 | 348328 | 53.1327 | 50.0633 |
| | 1000001 | 765618 | 53.3921 | 50.0256 |
| | 1000000 | 794893 | 53.0955 | 49.9547 |
| | 809174 | 615788 | 52.0070 | 49.9366 |
| | 700223 | 523954 | 52.2157 | 49.8893 |
| | 671985 | 495610 | 52.5703 | 49.9726 |
| 2012 | 637130 | 444918 | 52.3452 | 49.9407 |
| | 747131 | 514836 | 52.8000 | 49.9366 |
| | 712763 | 493688 | 52.6580 | 49.9526 |
| | 569122 | 383251 | 52.7004 | 49.9515 |
| | 700001 | 478278 | 52.7799 | 50.0068 |
| | 679502 | 474669 | 52.8559 | 50.0750 |
| | 575353 | 377511 | 52.8084 | 50.0244 |
| | 531883 | 333930 | 53.0978 | 49.9850 |
| | 645022 | 415363 | 52.9309 | 49.9931 |
| | 628876 | 392142 | 52.7412 | 49.9054 |
| | 549674 | 343278 | 52.8011 | 50.1270 |
| | 491495 | 302353 | 52.9646 | 50.0567 |
| 2013 | 557513 | 343712 | 52.9746 | 50.0249 |
| | 633520 | 393080 | 53.2057 | 49.9914 |
| | 494771 | 300216 | 53.2112 | 50.0783 |
| | 946883 | 668573 | 53.1128 | 49.9282 |
| | 807910 | 567905 | 52.3544 | 49.9424 |
| | 797716 | 559107 | 52.6101 | 49.9477 |
| | 694321 | 474354 | 52.4560 | 49.9961 |
| | 702068 | 483703 | 52.4935 | 49.9787 |
| | 686467 | 463690 | 52.6491 | 50.0126 |
| | 716444 | 485168 | 52.6984 | 50.0007 |
| | 486346 | 310329 | 52.8948 | 49.9045 |
| | 555246 | 358502 | 53.0118 | 50.0477 |
| **Mean** | | **449731** | **52.8409** | **49.9893** |
| **Skewness** | | | **2.8409** | **-0.0107** |
| **Total** | **23769443** | **16190304** | | |

**Wheat D1 (Finam)**

| date | original sample length | nonzero sample length | prob. of success | prob. of success (perm.) |
|---|---|---|---|---|
| 2000 | 242 | 237 | 51.6040 | 50.0641 |
| 2001 | 234 | 227 | 51.5893 | 49.4615 |
| 2002 | 251 | 241 | 50.4124 | 47.8731 |
| 2003 | 226 | 221 | 49.8124 | 48.4946 |
| 2004 | 251 | 244 | 50.2032 | 50.6407 |
| 2005 | 250 | 245 | 50.6920 | 49.2684 |
| 2006 | 250 | 247 | 50.8103 | 51.1082 |
| 2007 | 253 | 246 | 50.0188 | 48.6631 |
| 2008 | 252 | 249 | 52.1181 | 50.5674 |
| 2009 | 250 | 249 | 52.7725 | 50.8288 |
| 2010 | 264 | 258 | 51.3450 | 49.1852 |
| 2011 | 300 | 298 | 52.0048 | 52.2391 |
| 2012 | 303 | 301 | 47.7238 | 49.6256 |
| 2013 | 303 | 299 | 50.5233 | 49.7834 |
| **Mean** | | **254** | **50.8307** | **49.8431** |
| **Skewness** | | | **0.8307** | **-0.1569** |
| **Total** | **3629** | **3562** | | |

**Wheat H1 (Finam)**

| date | original sample length | nonzero sample length | prob. of success | prob. of success (perm.) |
|---|---|---|---|---|
| 2010 | 1137 | 1071 | 52.0946 | 49.8973 |
| 2011 | 4720 | 4472 | 50.2773 | 50.6511 |
| 2012 | 5163 | 4828 | 50.5005 | 49.4762 |
| 2013 | 4912 | 4446 | 50.8995 | 50.2864 |
| **Mean** | | **3704** | **50.9430** | **50.0778** |
| **Skewness** | | | **0.9430** | **0.0777** |
| **Total** | **15932** | **14817** | | |

**Wheat M30 (Finam)**

| date | original sample length | nonzero sample length | prob. of success | prob. of success (perm.) |
|---|---|---|---|---|
| 2010 | 2135 | 1964 | 51.1945 | 51.4924 |
| 2011 | 8854 | 8206 | 50.9650 | 50.3252 |
| 2012 | 9924 | 8952 | 50.5938 | 50.0470 |
| 2013 | 9443 | 8166 | 50.9674 | 50.3821 |
| **Mean** | | **6822** | **50.9302** | **50.5617** |
| **Skewness** | | | **0.9302** | **0.5617** |
| **Total** | **30356** | **27288** | | |

**Wheat M15 (Finam)**

| date | original sample length | nonzero sample length | prob. of success | prob. of success (perm.) |
|---|---|---|---|---|
| 2010 | 4088 | 3617 | 51.5964 | 49.9223 |
| 2011 | 16977 | 15040 | 51.0741 | 49.7479 |
| 2012 | 19301 | 16742 | 50.8285 | 50.2321 |
| 2013 | 18034 | 14948 | 51.5463 | 50.0721 |
| **Mean** | | **12587** | **51.2613** | **49.9936** |
| **Skewness** | | | **1.2613** | **-0.0064** |
| **Total** | **58400** | **50347** | | |

**Wheat M10 (Finam)**

| date | original sample length | nonzero sample length | prob. of success | prob. of success (perm.) |
|---|---|---|---|---|
| 2010 | 5945 | 5096 | 51.7749 | 49.5991 |
| 2011 | 24532 | 21384 | 51.6175 | 50.0720 |
| 2012 | 28054 | 23675 | 51.3780 | 49.9859 |
| 2013 | 25845 | 20887 | 51.5797 | 49.9143 |
| **Mean** | | **17761** | **51.5875** | **49.8928** |
| **Skewness** | | | **1.5875** | **-0.1072** |
| **Total** | **84376** | **71042** | | |

**Wheat M5 (Finam)**

| date | original sample length | nonzero sample length | prob. of success | prob. of success (perm.) |
|---|---|---|---|---|
| 2010 | 10847 | 9075 | 51.6938 | 49.6671 |
| 2011 | 44979 | 37726 | 51.6710 | 49.9404 |
| 2012 | 51712 | 42123 | 51.6230 | 49.8812 |
| 2013 | 46121 | 35872 | 52.0772 | 49.9952 |
| **Mean** | | **31199** | **51.7663** | **49.8710** |
| **Skewness** | | | **1.7663** | **-0.1290** |
| **Total** | **153659** | **124796** | | |

**Wheat M1 (Finam)**

| date | original sample length | nonzero sample length | prob. of success | prob. of success (perm.) |
|---|---|---|---|---|
| 2010 | 35605 | 27092 | 52.0170 | 49.9378 |
| 2011 | 148036 | 113728 | 51.9029 | 50.0023 |
| 2012 | 175312 | 129949 | 52.1527 | 50.0114 |
| 2013 | 148092 | 102593 | 52.5159 | 50.0243 |
| **Mean** | | **93341** | **52.1471** | **49.9940** |
| **Skewness** | | | **2.1471** | **-0.0061** |
| **Total** | **507045** | **373362** | | |

**Wheat ticks (Finam)**

| date | original sample length | nonzero sample length | prob. of success | prob. of success (perm.) |
|---|---|---|---|---|
| 2011 | 136311 | 82880 | 54.6947 | 49.9180 |
| | 118175 | 70901 | 54.0677 | 49.9104 |
| | 182464 | 116337 | 54.0078 | 49.8032 |
| | 102211 | 65598 | 53.6541 | 50.3003 |
| | 179500 | 120674 | 53.8774 | 49.8736 |
| | 132053 | 86661 | 54.0312 | 50.0369 |
| | 137312 | 87267 | 54.2645 | 50.1839 |
| | 104156 | 67874 | 53.3452 | 49.7834 |
| | 141171 | 86114 | 54.3913 | 49.9251 |
| | 145429 | 87140 | 53.9477 | 50.0017 |
| | 95156 | 53444 | 53.9753 | 50.1619 |
| | 96852 | 54837 | 54.2316 | 50.1887 |
| 2012 | 115581 | 62249 | 55.1102 | 50.0072 |
| | 88443 | 47500 | 54.3885 | 49.8789 |
| | 114894 | 62086 | 55.0575 | 49.8953 |
| | 77602 | 42766 | 54.2581 | 50.0538 |
| | 159408 | 90143 | 54.4119 | 50.0760 |
| | 117721 | 68688 | 53.4868 | 49.6783 |
| | 204678 | 129141 | 53.1632 | 50.0736 |
| | 116198 | 72278 | 52.6592 | 50.0477 |
| | 173018 | 100283 | 53.9413 | 49.8469 |
| | 166943 | 91436 | 54.3714 | 49.9442 |
| | 124722 | 66151 | 54.3076 | 50.0726 |
| | 107120 | 54429 | 54.7200 | 49.8898 |
| 2013 | 140474 | 71939 | 54.8111 | 49.6872 |
| | 110291 | 54962 | 54.2484 | 49.5788 |
| | 116846 | 58663 | 54.5437 | 50.1560 |
| | 87236 | 44960 | 53.6521 | 49.7509 |
| | 115981 | 57596 | 54.2121 | 49.9262 |
| | 81145 | 41155 | 54.6873 | 49.9623 |
| | 96072 | 47965 | 53.8696 | 50.1303 |
| | 63812 | 34180 | 53.6074 | 50.1126 |
| | 102451 | 48659 | 55.4985 | 49.6640 |
| | 108722 | 50454 | 55.0383 | 49.8315 |
| | 60696 | 27435 | 55.0939 | 50.1094 |
| | 72930 | 32483 | 55.7971 | 49.7937 |
| **Mean** | | **67704** | **54.2618** | **49.9515** |
| **Skewness** | | | **4.2618** | **-0.0485** |
| **Total** | **4293774** | **2437328** | | |

**Google D1 (Finam)**

| date | original sample length | nonzero sample length | prob. of success | prob. of success (perm.) |
|---|---|---|---|---|
| 2004 | 92 | 92 | 49.6976 | 51.2776 |
| 2005 | 250 | 250 | 50.3697 | 49.9471 |
| 2006 | 249 | 249 | 51.7213 | 51.6003 |
| 2007 | 250 | 250 | 50.3455 | 48.7226 |
| 2008 | 273 | 273 | 51.4418 | 51.4463 |
| 2009 | 282 | 267 | 46.7975 | 50.5436 |
| 2010 | 250 | 250 | 50.1916 | 51.8110 |
| 2011 | 225 | 225 | 48.9599 | 47.3084 |
| 2012 | 248 | 248 | 51.1114 | 48.7578 |
| 2013 | 250 | 249 | 48.9688 | 50.5284 |
| **Mean** | | **235** | **49.9605** | **50.1943** |
| **Skewness** | | | **-0.0395** | **0.1943** |
| **Total** | **2369** | **2353** | | |

**Google H1 (Finam)**

| date | original sample length | nonzero sample length | prob. of success | prob. of success (perm.) |
|---|---|---|---|---|
| 2009 | 1191 | 1184 | 50.8093 | 49.8780 |
| 2010 | 2011 | 1995 | 50.7183 | 50.0588 |
| 2011 | 1579 | 1573 | 50.3980 | 49.6250 |
| 2012 | 1719 | 1712 | 51.0447 | 50.8222 |
| 2013 | 1737 | 1736 | 50.2782 | 49.4384 |
| **Mean** | | **1640** | **50.6497** | **49.9645** |
| **Skewness** | | | **0.6497** | **-0.0355** |
| **Total** | **8237** | **8200** | | |

**Google M30 (Finam)**

| date | original sample length | nonzero sample length | prob. of success | prob. of success (perm.) |
|---|---|---|---|---|
| 2010 | 3521 | 3507 | 50.3370 | 50.0143 |
| 2011 | 2895 | 2880 | 50.3052 | 49.7049 |
| 2012 | 3188 | 3177 | 50.7378 | 49.9078 |
| 2013 | 3217 | 3208 | 50.8507 | 50.2142 |
| **Mean** | | **3193** | **50.5577** | **49.9603** |
| **Skewness** | | | **0.5577** | **-0.0397** |
| **Total** | **12821** | **12772** | | |

**Google M15 (Finam)**

| date | original sample length | nonzero sample length | prob. of success | prob. of success (perm.) |
|---|---|---|---|---|
| 2010 | 6788 | 6742 | 50.8484 | 50.6939 |
| 2011 | 5733 | 5712 | 50.8480 | 49.6549 |
| 2012 | 6373 | 6341 | 50.3827 | 50.2329 |
| 2013 | 6424 | 6389 | 50.6676 | 49.9429 |
| **Mean** | | **6296** | **50.6867** | **50.1312** |
| **Skewness** | | | **0.6867** | **0.1312** |
| **Total** | **25318** | **25184** | | |

**Google M10 (Finam)**

| date | original sample length | nonzero sample length | prob. of success | prob. of success (perm.) |
|---|---|---|---|---|
| 2010 | 10054 | 9982 | 51.2534 | 50.1563 |
| 2011 | 8555 | 8500 | 50.7215 | 49.8256 |
| 2012 | 9551 | 9490 | 50.7297 | 49.9248 |
| 2013 | 9613 | 9562 | 50.5913 | 50.1028 |
| **Mean** | | **9384** | **50.8240** | **50.0024** |
| **Skewness** | | | **0.8240** | **0.0024** |
| **Total** | **37773** | **37534** | | |

**Google M5 (Finam)**

| date | original sample length | nonzero sample length | prob. of success | prob. of success (perm.) |
|---|---|---|---|---|
| 2010 | 19848 | 19685 | 51.0983 | 49.9101 |
| 2011 | 16962 | 16807 | 50.8531 | 50.0788 |
| 2012 | 18941 | 18780 | 51.0152 | 50.0218 |
| 2013 | 18755 | 18599 | 50.8629 | 49.8103 |
| **Mean** | | **18468** | **50.9574** | **49.9553** |
| **Skewness** | | | **0.9574** | **-0.0448** |
| **Total** | **74506** | **73871** | | |

**Google M1 (Finam)**

| date | original sample length | nonzero sample length | prob. of success | prob. of success (perm.) |
|---|---|---|---|---|
| 2010 | 97723 | 95584 | 51.2476 | 50.0571 |
| 2011 | 72637 | 71107 | 50.6930 | 50.0410 |
| 2012 | 73510 | 72009 | 50.4893 | 50.0346 |
| 2013 | 63699 | 62606 | 50.4095 | 50.0561 |
| **Mean** | | **75327** | **50.7099** | **50.0472** |
| **Skewness** | | | **0.7098** | **0.0472** |
| **Total** | **307569** | **301306** | | |

**Google ticks (Finam)**

| date | original sample length | nonzero sample length | prob. of success | prob. of success (perm.) |
|---|---|---|---|---|
| 2011 | 38496 | 24897 | 50.1567 | 49.5501 |
| | 48848 | 34140 | 49.2164 | 49.9692 |
| | 42128 | 30397 | 49.2022 | 49.5608 |
| | 35248 | 25442 | 48.8051 | 50.3125 |
| 2012 | 46936 | 32399 | 49.3919 | 50.3843 |
| | 31604 | 21835 | 48.2253 | 50.0893 |
| | 29829 | 21740 | 48.7235 | 49.7424 |
| | 33682 | 24213 | 48.6949 | 50.1053 |
| | 37329 | 26546 | 48.4649 | 49.6101 |
| | 31098 | 21231 | 48.7329 | 50.0424 |
| | 33723 | 24070 | 49.1608 | 50.4383 |
| | 27376 | 21221 | 48.7960 | 50.4265 |
| | 35455 | 27967 | 48.9362 | 49.7944 |
| | 42670 | 33002 | 48.1728 | 49.8046 |
| | 29230 | 22838 | 48.8878 | 49.9869 |
| | 22521 | 17762 | 48.9612 | 49.4342 |
| 2013 | 28787 | 22226 | 48.9922 | 50.0900 |
| | 27209 | 20755 | 48.9376 | 49.6459 |
| | 25548 | 18838 | 49.0046 | 49.5036 |
| | 31537 | 22084 | 48.7660 | 49.9955 |
| | 27426 | 18965 | 48.8110 | 49.8998 |
| | 33314 | 22611 | 48.9231 | 50.4246 |
| | 30765 | 21383 | 48.8121 | 49.8667 |
| | 15061 | 12479 | 49.7195 | 49.8477 |
| | 15786 | 12783 | 49.2920 | 49.7770 |
| | 28738 | 22777 | 48.4413 | 50.3512 |
| | 12112 | 9595 | 48.5303 | 49.8385 |
| | 13676 | 10799 | 48.4256 | 50.3010 |
| **Mean** | | **22321** | **48.8994** | **49.9569** |
| **Skewness** | | | **-1.1006** | **-0.0431** |
| **Total** | **856132** | **624995** | | |

**Apple D1 (Finam)**

| date | original sample length | nonzero sample length | prob. of success | prob. of success (perm.) |
|---|---|---|---|---|
| 2000 | 251 | 246 | 53.1493 | 50.0413 |
| 2001 | 247 | 240 | 53.0089 | 50.8657 |
| 2002 | 251 | 245 | 49.0528 | 51.5300 |
| 2003 | 251 | 246 | 50.7284 | 50.6311 |
| 2004 | 250 | 245 | 50.3969 | 49.7335 |
| 2005 | 250 | 249 | 50.1026 | 49.6153 |
| 2006 | 249 | 248 | 50.3056 | 50.9251 |
| 2007 | 250 | 250 | 50.9829 | 50.1594 |
| 2008 | 273 | 271 | 53.9314 | 49.1510 |
| 2009 | 280 | 266 | 47.3347 | 49.6453 |
| 2010 | 250 | 250 | 48.3537 | 51.8148 |
| 2011 | 225 | 225 | 47.0934 | 51.2241 |
| 2012 | 248 | 247 | 49.4806 | 50.9871 |
| 2013 | 250 | 250 | 51.6349 | 49.5370 |
| **Mean** | | **248** | **50.3969** | **50.4186** |
| **Skewness** | | | **0.3969** | **0.4186** |
| **Total** | **3525** | **3478** | | |

**Apple H1 (Finam)**

| date | original sample length | nonzero sample length | prob. of success | prob. of success (perm.) |
|---|---|---|---|---|
| 2009 | 1194 | 1182 | 49.9262 | 49.9832 |
| 2010 | 2012 | 1982 | 50.3208 | 50.3821 |
| 2011 | 1581 | 1573 | 50.8903 | 49.3774 |
| 2012 | 1722 | 1718 | 49.0030 | 50.1232 |
| 2013 | 1737 | 1733 | 50.4123 | 48.4724 |
| **Mean** | | **1638** | **50.1105** | **49.6677** |
| **Skewness** | | | **0.1105** | **-0.3323** |
| **Total** | **8246** | **8188** | | |

**Apple M30 (Finam)**

| date | original sample length | nonzero sample length | prob. of success | prob. of success (perm.) |
|---|---|---|---|---|
| 2010 | 3524 | 3476 | 50.1145 | 49.9692 |
| 2011 | 2900 | 2881 | 50.5078 | 50.0774 |
| 2012 | 3191 | 3182 | 50.4705 | 50.7316 |
| 2013 | 3218 | 3203 | 50.4589 | 50.5548 |
| **Mean** | | **3186** | **50.3879** | **50.3333** |
| **Skewness** | | | **0.3879** | **0.3333** |
| **Total** | **12833** | **12742** | | |

**Apple M15 (Finam)**

| date | original sample length | nonzero sample length | prob. of success | prob. of success (perm.) |
|---|---|---|---|---|
| 2010 | 6796 | 6703 | 50.4276 | 50.1013 |
| 2011 | 5738 | 5686 | 50.6585 | 49.6231 |
| 2012 | 6374 | 6336 | 50.4159 | 50.0240 |
| 2013 | 6428 | 6378 | 50.8921 | 49.9704 |
| **Mean** | | **6276** | **50.5985** | **49.9297** |
| **Skewness** | | | **0.5985** | **-0.0703** |
| **Total** | **25336** | **25103** | | |

**Apple M10 (Finam)**

| date | original sample length | nonzero sample length | prob. of success | prob. of success (perm.) |
|---|---|---|---|---|
| 2010 | 10064 | 9919 | 50.7706 | 50.0147 |
| 2011 | 8562 | 8492 | 50.9583 | 50.2580 |
| 2012 | 9552 | 9487 | 50.6048 | 50.3689 |
| 2013 | 9631 | 9566 | 50.8619 | 49.9322 |
| **Mean** | | **9366** | **50.7989** | **50.1435** |
| **Skewness** | | | **0.7989** | **0.1434** |
| **Total** | **37809** | **37464** | | |

**Apple M5 (Finam)**

| date | original sample length | nonzero sample length | prob. of success | prob. of success (perm.) |
|---|---|---|---|---|
| 2010 | 19863 | 19481 | 50.8478 | 49.9317 |
| 2011 | 17034 | 16824 | 50.7103 | 50.0055 |
| 2012 | 19088 | 18953 | 50.6097 | 50.2969 |
| 2013 | 19255 | 19087 | 50.7827 | 49.9528 |
| **Mean** | | **18586** | **50.7376** | **50.0467** |
| **Skewness** | | | **0.7376** | **0.0467** |
| **Total** | **75240** | **74345** | | |

**Apple M1 (Finam)**

| date | original sample length | nonzero sample length | prob. of success | prob. of success (perm.) |
|---|---|---|---|---|
| 2010 | 98160 | 94449 | 50.6973 | 49.9177 |
| 2011 | 84062 | 81719 | 50.7288 | 50.0725 |
| 2012 | 94722 | 92999 | 50.7706 | 49.9759 |
| 2013 | 93711 | 91672 | 50.8827 | 49.9708 |
| **Mean** | | **90210** | **50.7699** | **49.9842** |
| **Skewness** | | | **0.7698** | **-0.0158** |
| **Total** | **370655** | **360839** | | |

**Apple ticks (Finam)**

| date | original sample length | nonzero sample length | prob. of success | prob. of success (perm.) |
|---|---|---|---|---|
| 2011 | 270463 | 166822 | 48.7786 | 50.0923 |
| | 319909 | 198259 | 48.7463 | 50.0393 |
| | 234604 | 146933 | 48.8069 | 50.0684 |
| | 147845 | 93583 | 48.5366 | 50.4119 |
| 2012 | 153719 | 91966 | 49.3715 | 50.0663 |
| | 236809 | 151967 | 48.9136 | 50.3465 |
| | 272221 | 183861 | 48.7708 | 50.0495 |
| | 258969 | 174405 | 48.7194 | 50.0401 |
| | 222662 | 152415 | 48.4716 | 50.1512 |
| | 160613 | 110776 | 48.7411 | 49.9756 |
| | 197174 | 136997 | 48.5970 | 50.0259 |
| | 163015 | 112350 | 48.5376 | 50.1113 |
| | 199025 | 138613 | 48.3270 | 49.9776 |
| | 244923 | 170684 | 48.4178 | 49.8942 |
| | 225577 | 161534 | 48.1348 | 50.0446 |
| | 211445 | 149871 | 48.2064 | 49.9223 |
| 2013 | 204995 | 138982 | 48.5480 | 50.1076 |
| | 170129 | 115446 | 48.5703 | 49.9437 |
| | 164109 | 108272 | 48.2604 | 49.9801 |
| | 179751 | 115610 | 48.1667 | 49.9905 |
| | 154164 | 102352 | 48.2086 | 50.3312 |
| | 133561 | 92132 | 48.2243 | 49.7031 |
| | 120276 | 84633 | 48.4368 | 49.8452 |
| | 124559 | 87907 | 48.5758 | 50.1803 |
| | 133516 | 92247 | 48.5891 | 50.0927 |
| | 123408 | 86080 | 48.2795 | 49.6974 |
| | 82012 | 56612 | 48.3334 | 49.8225 |
| | 99078 | 70077 | 48.3525 | 50.0885 |
| **Mean** | | **124692** | **48.5222** | **50.0357** |
| **Skewness** | | | **-1.4778** | **0.0357** |
| **Total** | **5208531** | **3491386** | | |